\documentclass[twocolumn,trackchanges]{aastex6}
\RequirePackage{lineno}
\usepackage{amsmath}
\usepackage{amssymb}
\usepackage{amsthm}
\usepackage{natbib,aasdefs,url,bm}
\usepackage{array}
\usepackage{float}
\usepackage{graphicx}
\usepackage{subfigure}
\usepackage{color}
\usepackage{threeparttable} 
\usepackage{CJK}
\usepackage{lineno}

\newcounter{ichi}
\newcounter{ni}
\newcounter{san}
\newcounter{yon}
\def\be{\begin{equation}}
\def\ee{\end{equation}}
\def\ba{\begin{eqnarray}}
\def\ea{\end{eqnarray}}

\newcommand{\re}{\color{black}}

\def\aap{\ {A\&A}\ }

\def\apjl{\ {ApJL}\ }
\def\apjs{\ {ApJS}\ }

\shorttitle{Reverse shock origin of VHE gamma-ray emission from GRB 221009A}
\shortauthors{Zhang et al.}

\begin{document}
\begin{CJK*}{UTF8}{gbsn}

\title{External Inverse-Compton and Proton Synchrotron Emission from the Reverse Shock as the Origin of VHE Gamma-Rays from the Hyper-Bright GRB 221009A}

\author{B. Theodore Zhang (张兵)\altaffilmark{1}}

\author{Kohta Murase\altaffilmark{2,3,4,5,1}}

\author{Kunihito Ioka\altaffilmark{1}}

\author{Deheng Song\altaffilmark{1}}

\author{Chengchao Yuan (袁成超)\altaffilmark{6}}

\author{P\'eter M\'esz\'aros\altaffilmark{2,3,4}}

\altaffiltext{1}{Center for Gravitational Physics and Quantum Information, Yukawa Institute for Theoretical Physics, Kyoto University, Kyoto, Kyoto 606-8502, Japan}
\altaffiltext{2}{Department of Physics, The Pennsylvania State University, University Park, PA 16802, USA}
\altaffiltext{3}{Department of Astronomy \& Astrophysics, The Pennsylvania State University, University Park, PA 16802, USA}
\altaffiltext{4}{Center for Multimessenger Astrophysics, Institute for Gravitation and the Cosmos, The Pennsylvania State University, University Park, PA 16802, USA}
\altaffiltext{5}{School of Natural Sciences, Institute for Advanced Study, Princeton, NJ 08540, USA}
\altaffiltext{6}{Deutsches Elektronen-Synchrotron DESY, Platanenallee 6, 15738 Zeuthen, Germany}

\begin{abstract}
The detection of the hyper-bright gamma-ray burst (GRB) 221009A enables us to explore the nature of the GRB emission and the origin of very-high-energy (VHE) gamma-rays. 
We analyze the {\it Fermi}-LAT data of this burst and investigate the GeV-TeV emission in the framework of the external reverse shock model. 
We show that the early $\sim1-10$ GeV emission can be explained by the external inverse-Compton mechanism via upscattering MeV gamma-rays by electrons accelerated at the reverse shock, in addition to the synchrotron self-Compton component. 
The predicted early optical flux could have been brighter than that of the naked-eye GRB 080319B.
We also show that proton synchrotron emission from accelerated ultra-high-energy cosmic rays (UHECRs) is detectable, and could potentially explain $\gtrsim \rm TeV$ photons detected by LHAASO or constrain the UHECR acceleration mechanism.
Our model suggests that the detection of $\mathcal{O}(10\rm~TeV)$ photons with energies up to $\sim18$ TeV is possible for reasonable models of the extragalactic background light without invoking new physics, and predicts anti-correlations between MeV photons and TeV photons, which can be tested with the LHAASO data.
\end{abstract}

\section{Introduction}
Gamma-ray bursts (GRBs) are among the most luminous explosions in the Universe~\citep{meszaros_gamma-ray_2006,Kumar:2014upa}.
In 2019, the detection of two TeV bursts, GRB 190114C~\citep{magic_collaboration_teraelectronvolt_2019, magic_collaboration_observation_2019} and GRB 180720B~\citep{abdalla_very-high-energy_2019}, has opened a new window in the VHE ($\gtrsim 0.1\rm~TeV$) band for studying GRBs, providing us with new opportunities to investigate the nature of GRBs~\citep[see][for reviews]{miceli_gamma-ray_2022,Gill:2022erf}.

On October 9 2022, GRB 221009A was triggered by the Fermi Gamma-Ray Burst Monitor (GBM) at $T_0$ = 13:16:59.99 UT~\citep{veres_grb_2022}. 
The Swift Burst Alert Telescope (BAT) also triggered GRB 221009A around one-hour later~\citep{dichiara_swift_2022}.
GRB 221009A is an extraordinarily bright and energetic GRB with isotropic-equivalent energy $\mathcal{E}_{k} \sim 3 \times 10^{55}\rm~erg$ for a radiative efficiency of 10\%~\citep{frederiks_konus-wind_2022} at a redshift $z = 0.15$~\citep{de_ugarte_postigo_grb_2022}.
The \textit{Fermi}-LAT reported the detection of $> 100\rm~MeV$ gamma-rays with the maximum photon energy reaching 99.3 GeV~\citep{pillera_grb_2022}.
Remarkably, it was reported by the Large High Altitude Air Shower Observatory (LHAASO) that there are more than 5000 gamma-rays with energy beyond 500 GeV from GRB 221009A detected, and the highest-energy gamma-ray energy reaches 18 TeV~\citep{huang_lhaaso_2022}.

The production of VHE gamma-rays from GRB up to $\sim \rm TeV$ has been widely discussed in the standard afterglow model via synchrotron self-Compton (SSC) process~\citep[e.g.,][]{Meszaros:1994sd, Dermer:1999eh, Sari:2000zp, Zhang:2001az, ren_very_2022} or external inverse-Compton (EIC) process~\citep[e.g.,][]{Wang:2006eq, Murase+10,  toma_photosphere-internal_2011, murase_implications_2011, He:2011aa, Veres:2013dea, Kimura:2019fae, zhang_external_2021, zhang_external_sGRB_2021}.
Proton synchrotron emission has been proposed as one alternative mechanism to generate VHE gamma-rays from GRBs~\citep[e.g.,][]{totani_very_1998, Zhang:2001az, murase_high-energy_2008, asano_prompt_2009, isravel_proton_2022}, which usually requires protons to be accelerated to the ultra-high-energy (UHE) range.
The proton synchrotron emission has an advantage in the generation of $\gtrsim 10\rm~TeV$ gamma-rays, which is usually difficult for SSC and EIC processes due to the limitation of the Klein-Nishina effect.
Motivated by the detection of $\mathcal{O}(10\rm~TeV)$ gamma-rays from GRB 221009A by the LHAASO observatory, it is reasonable to study the proton synchrotron radiation process of GRB 221009A in detail.

It has been proposed that the acceleration of UHECRs is possible in the internal shock model~\citep[e.g.,][]{waxman_cosmological_1995, vietri_coronal_1996} or reverse shock model of GRBs~\citep[e.g.,][]{waxman_neutrino_2000, murase_high_2007, murase_high-energy_2008, zhang_low-luminosity_2018},
while the acceleration of UHECRs in the forward shock region via the diffusive shock acceleration mechanism is difficult due to the low magnetic field strength of the external medium~\citep{gallant_ultra-high-energy_1999,murase_high-energy_2008,SKL15}. 
GRB internal shocks may occur at much smaller radii, where the escape of $\mathcal{O}(10\rm~TeV)$ gamma-rays is difficult unless the Lorentz factor is extremely large~\citep[e.g.,][]{murase_neutrinos_2022}. 
One of the possibilities is the UHECR proton synchrotron emission process in the reverse shock model.
The observed prompt emission in the MeV band indicates that GRB 221009A is a long-lasting GRB with $T_{\rm 90}$ of at least $\sim 600$ seconds, implying a thick ejecta shell case where the long-lasting reverse shock can be expected.
Also, during the early phase of the reverse shock, the extraordinarily bright prompt photons can be upscattered by the high-energy electrons accelerated in the reverse shock region to produce high-energy gamma-rays which may be related to \textit{Fermi}-LAT observations~\citep[e.g.,][]{Beloborodov05}.

Throughout this work, we use $Q_x/Q=10^{x}$ in CGS units.

\section{GeV-TeV gamma-rays from a reverse shock model}\label{sec:mod}

\begin{figure}[th]
\includegraphics[width=\linewidth]{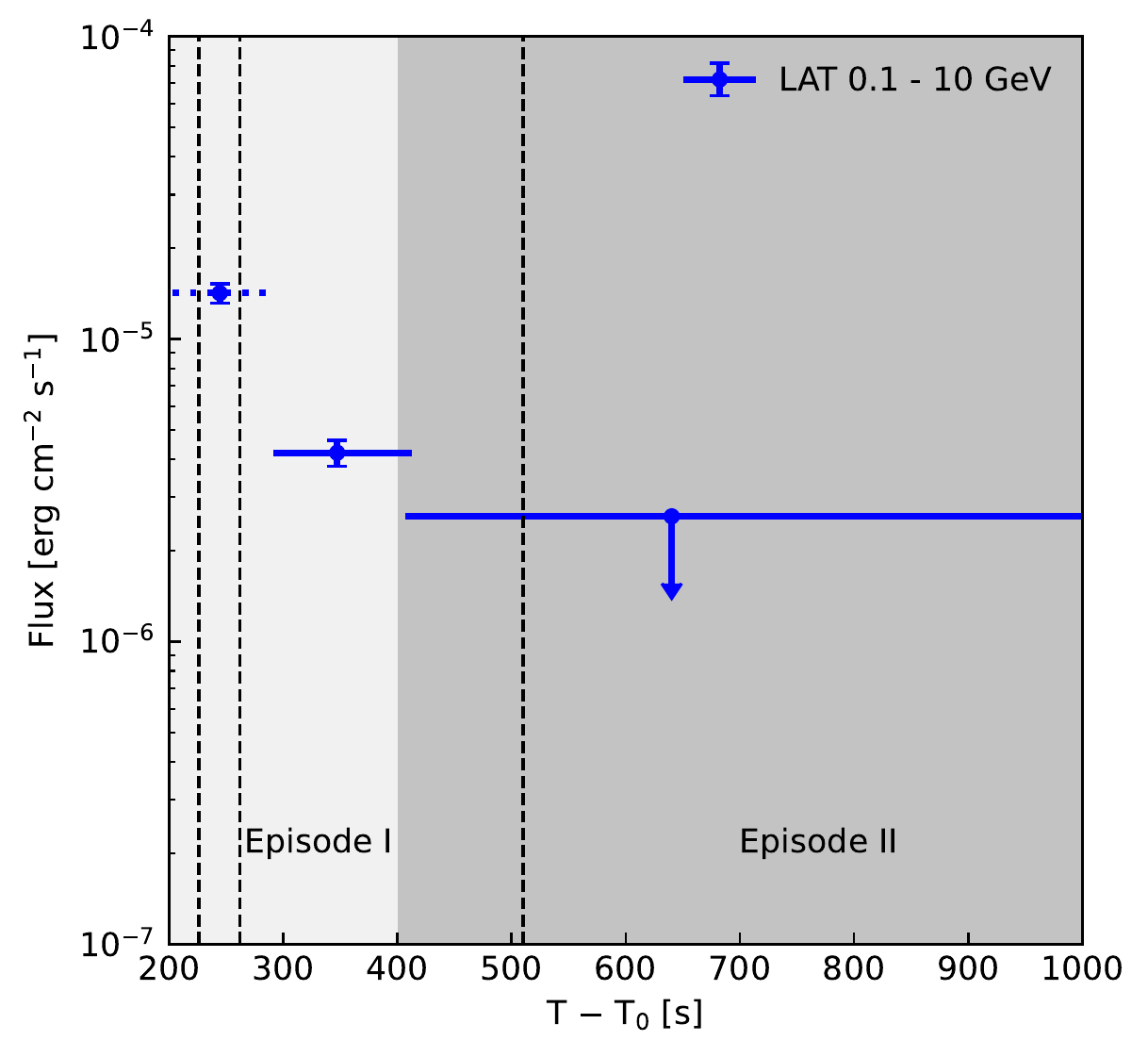}
\caption{The flux light curve observed by the \textit{Fermi}-LAT between $T_0$ + 203~s and $T_0$ + 1000~s. The first time interval overlaps with the LAT Bad Time Interval due to photon pile-up and therefore is only included for reference purposes. The black-dashed lines indicate the three peak times of the prompt emission observed by the Konus-Wind.
\label{fig:lc}
}
\end{figure}

We analyze the \textit{Fermi}-LAT data collected between 203~s and 1000~s after the Fermi GBM trigger time of $T_0$ = 13:16:59.99 UT~\citep{veres_grb_2022, pillera_grb_2022}.
Our region of interest (ROI) is a 10$^\circ$ by 10$^\circ$ region centered around the location of the GRB as reported by \citealt{pillera_grb_2022} (RA = 288.28$^\circ$, DEC = 19.49$^\circ$). We select the events from the \texttt{P8R3\_TRANSIENT020} class with energies between 0.1 GeV and 10 GeV. We consider 3 time intervals after $T_0$: 203~s - 294~s, 294~s - 410~s, and 410~s - 1000~s. In each time interval, we perform unbinned likelihood analysis by varying a point source placed at the center of the ROI. We also include the Galactic interstellar emission model \texttt{gll\_iem\_v07.fits} and the isotropic diffuse emission model \texttt{iso\_P8R3\_TRANSIENT020\_V3\_v1.txt}, and their normalizations are allowed to vary freely. The first time interval overlaps with the LAT {\re Bad} Time Interval due to photon pile-up \citep{omodei_grb_2022_high, omodei_grb_2022}. We only include it for reference purposes. The LAT photons were detected in the time interval $T_0$ + 294s to $T_0$ + 410~s, and the integrated photon flux between 0.1 GeV and 10 GeV is $(4.0 \pm 0.4)\times 10^{-3}$ ph cm$^{-2}$~s$^{-1}$ with a photon index of $1.76 \pm 0.09$. We do not find photons from the direction of the GRB during the time interval $T_0$ + 410~s to $T_0$ + 1000~s. The 95\% upper limit of the flux is $3.5\times 10^{-3}$ ph cm$^{-2}$ s$^{-1}$. The flux light curve is shown in Fig.~\ref{fig:lc}. 

In Fig.~\ref{fig:lc}, we also show the peak time of the observed pulse of GRB prompt emission detected by Konus-Wind at 13:17:01.648 UT~\citep{frederiks_konus-wind_2022}.
We can see there was active prompt emission during the \textit{Fermi}-LAT observations within the time interval $T - T_0 \sim 200 - 400\rm~s$, while the prompt emission becomes weaker at later times up to {\re $\sim 600\rm~s$, which may also be regarded as the flare phase}~\citep{Frederiks:2023bxg, Insight-HXMT:2023aof}. 
Considering the long-lasting prompt emission of GRB 221009A, in this work, we assume that the GRB ejecta is thick with duration time {\re $600\rm~s$}~\citep[e.g.,][]{ioka_variabilities_2005}.

We consider two episodes, where the first episode (Episode I) is strongly affected by the active prompt emission phase, while the second episode (Episode II) is much less affected by the prompt emission until the reverse shock finishes crossing the ejecta. 
However, the detailed modeling of the dynamical evolution of the reverse shock can be complicated~\citep{sari_hydrodynamic_1995, sari_hydrodynamics_1997}.
For the purpose of this work, we analytically derive 
the characteristic radius and time of the reverse shock at the shock crossing time $t_\times$~\citep{panaitescu_analysis_2004, murase_high_2007}, where the reverse shock finishes crossing the ejecta.

When the ultrarelativistic thick ejecta propagates into the external medium, two types of shocks are formed: a reverse shock (RS), which propagates back into the ejecta shell increasing the internal energy, and a forward shock (FS), which propagates into the external medium and energizes the swept-up matter.
The typical crossing time $t_\times$ for the RS to complete its crossing through the ejecta depends on the width of the ultrarelativistic ejecta shell $\Delta$ that represents the geometrical thickness measured in the stellar frame.
In the thick ejecta shell regime, the ejecta width can be estimated as {\re $\Delta \approx c \delta T \simeq 1.9 \times 10^{13}~\delta T_{2.8}\rm~cm$}, where ${\re \delta}T$ is the duration of the GRB ejecta released by their source measured in the GRB frame.
Then the crossing radius of the reverse shock is {\re $r_\times \simeq 3.8 \times 10^{17}~\mathcal{E}_{k, 55}^{1/4} n^{-1/4} \delta T_{2.8}^{1/4} \rm~cm$} where $\mathcal{E}_k$ is the isotropic-equivalent kinetic energy and $n$ is the constant density of the external medium. The crossing time is {\re $t_\times (\delta T) \approx 0.71 \delta T (1 + z) \simeq 540~{\re \delta}T_{2.8} \rm~s$}, and hereafter the dependence on $z$ is neglected for simplicity. 
The Lorentz factor of the shocked ejecta at the crossing radii $r_\times$ can be written as {\re $\Gamma_\times \simeq 83~\mathcal{E}_{k, 55}^{1/8} n^{-1/8} \delta T_{2.8}^{-3/8}$} and {\re $\Gamma_{\rm rel} \approx 0.5(\Gamma_\times / \Gamma_0 + \Gamma_0 / \Gamma_\times) \simeq 1.7 \Gamma_{0, 2.4}$} when measured in the stellar frame and in the frame of unshocked ejecta, respectively. Here $\Gamma_0$ is the initial Lorentz factor of GRB ejecta. The reverse shock is relativistic before it crosses the shell. The contribution to the shell width due to the velocity spread can be estimated as {\re $r_\times / 2\Gamma_0^2 \simeq 3.1\times 10^{12} \mathcal{E}_{k, 55}^{1/4} n^{-1/4} \delta T_{2.8}^{1/4} \Gamma_{0,2.4}^{-2}\rm~cm < \Delta$} as expected in the thick ejecta shell regime~\citep{sari_hydrodynamic_1995}.

For the earlier evolution phase of the reverse shock, especially for Episode I, we assume that only a fraction of the total ejecta energy $\mathcal{E}_{k, \rm I} \approx ({\re \delta T_{\rm I} / \delta}T) \mathcal{E}_k$ carried by the outer edge of the ejecta shell is transferred to the external medium during ${\re t_{\times,{\rm I}}\equiv t_\times (\delta T_{\rm I}) < t_\times}$. Thus, we can use the same method {\re as above} to estimate the characteristic radius and Lorentz factor of the shocked ejecta after the reverse shock crosses.
{\re For Episode II, we adopt the same assumptions and use the same method, except that $\delta T_{\rm II} \sim \delta T$ and $\mathcal{E}_{k, \rm II} \sim \mathcal{E}_{k}$.}

Now we proceed to derive the emission properties of the shocked ejecta at the shock crossing radius for both episodes. 
{\re Note that for Episode I, the physical quantities in the following equations are normalized to the numerical values closer to those expected for Episode II.}
For Episode II, the magnetic field strength in the comoving frame is estimated as {\re $B_\times = [32 \pi \epsilon_B n_{\rm ej} m_p c^2 (\Gamma_{\rm rel} - 1) (\Gamma_{\rm rel} + 3/4)]^{1/2} \simeq 8.6{\rm~G} \epsilon_{B, -1}^{1/2} \mathcal{E}_{k, 55}^{1/4} \Gamma_{0,2.4}^{-1} \delta T_{2.8}^{-3/4} n^{1/4} (g(\Gamma_{\rm rel}) / 1.7)^{1/2}$}, where $\epsilon_B$ is the energy fraction of internal energy that is converted into the magnetic energy, the proton number density of the unshocked ejecta is {\re $n_{\rm ej} \simeq 3.2\times 10^3~\mathcal{E}_{k, 55}^{1/2} \Gamma_{0,2.4}^{-2} \delta T_{2.8}^{-3/2} n^{1/2} \rm~cm^{-3}$} and $g(\Gamma_{\rm rel}) \equiv (\Gamma_{\rm rel} - 1) (\Gamma_{\rm rel} + 3/4)$. 
The total number of electrons {\re energized} by the reverse shock is {\re $N_e^r = \mathcal{E}_k / \Gamma_0 m_p c^2 \simeq 2.8 \times 10^{55}~\mathcal{E}_{k, 55} \Gamma_{0,2.4}^{-1}$}. The minimum electron Lorentz factor after shock acceleration can be estimated as {\re $\gamma_m \approx (\epsilon_e / f_e) [(s_e - 2) / (s_e - 1)] (m_p / m_e) (\Gamma_{\rm rel} - 1)  \simeq 4.2 \times 10^2 \epsilon_{e, -1} f_{e, -1}^{-1}$} for $s_e = 2.6$ and {\re $\Gamma_{\rm rel} = 1.7$}, where the energy fraction $\epsilon_e$ of the post-shock internal energy is converted into electron non-thermal energy, $f_e$ is the number fraction of accelerated electrons and $s_e$ is the electron spectral index.
{\re The main difference between the treatment of Episode I and Episode II are the values adopted for $\epsilon_B$ and $\epsilon_e$ (see Table I).}
We then derive the steady-state electron energy distribution considering various cooling processes, including adiabatic cooling, synchrotron cooling, SSC cooling, and EIC cooling, using the iteration method described in~\cite{murase_implications_2011, zhang_external_2021}.
The comoving frame non-thermal proton energy density is determined by $U_p \approx \epsilon_p \mathcal{E}_{k} / (4\pi r_\times^2 \Gamma_\times^2 \Delta$), where $\epsilon_{p}$ is the fraction of downstream energy transferred to the non-thermal protons.
The comoving frame minimum proton energy is $\varepsilon_{p, \rm min} \approx \Gamma_{\rm rel} m_p c^2$.
The maximum proton energy achieved under the confinement condition $t_{\rm acc} < t_{\rm dyn}$ is {\re $E_{\rm max, dyn} \approx \eta^{-1} e B_\times r_\times \simeq 1.0 \times 10^{21}~{\rm~eV}~\eta^{-1} \mathcal{E}_{k, 55}^{1/2} \Gamma_{0,2.4}^{-1} \delta T_{2.8}^{-1/2} \epsilon_{B, -1}^{1/2} (g(\Gamma_{\rm rel}) / 1.7)^{1/2}$}.
{\re Here, we define the acceleration timescale as $t_{\rm acc} = \eta t_L$, where $t_L$ is the Larmor time, and $\eta$ is a coefficient which is $\sim$a~few in the Bohm limit ~\citep[e.g.,][]{SKL15}.}
The production of UHECRs at the GRB reverse shock is possible~\citep[e.g.,][]{murase_high-energy_2008, zhang_low-luminosity_2018}, where the maximum proton energy can be limited by various cooling processes, e.g., for synchrotron cooling we have {\re $E_{\rm max, syn} \simeq 5.5 \times 10^{20}~{\rm~eV}~Z^{-3/2} \eta^{-1/2} \epsilon_{B, -1}^{-1/4} \Gamma_{0, 2.4}^{1/2} n^{-1/4} (g(\Gamma_{\rm rel}) / 1.7)^{-1/4}$}. 
We also include the effect of photomeson cooling on the maximum energy of protons which depends on the energy density of the target photon fields.
The effect of overlapping of the prompt emission in the reverse shocked region is considered, where the comoving prompt photon energy density measured in the reverse shocked region is $U_{\rm GRB\gamma} \approx L_{\rm GRB\gamma}^{\rm iso} / 4\pi r_\times^2 \Gamma_\times^2 c$. The Band function is used for modeling the energy spectra of the prompt emission, where the low-energy power law index is $\alpha = 1.1$, the high-energy power law index is $\beta = 2.6$ and the observed peak energy is $E_{\rm pk} = 1\rm~MeV$~\citep{frederiks_konus-wind_2022}.

\section{Results}\label{sec:res}
\begin{figure}[th]
\includegraphics[width=\linewidth]{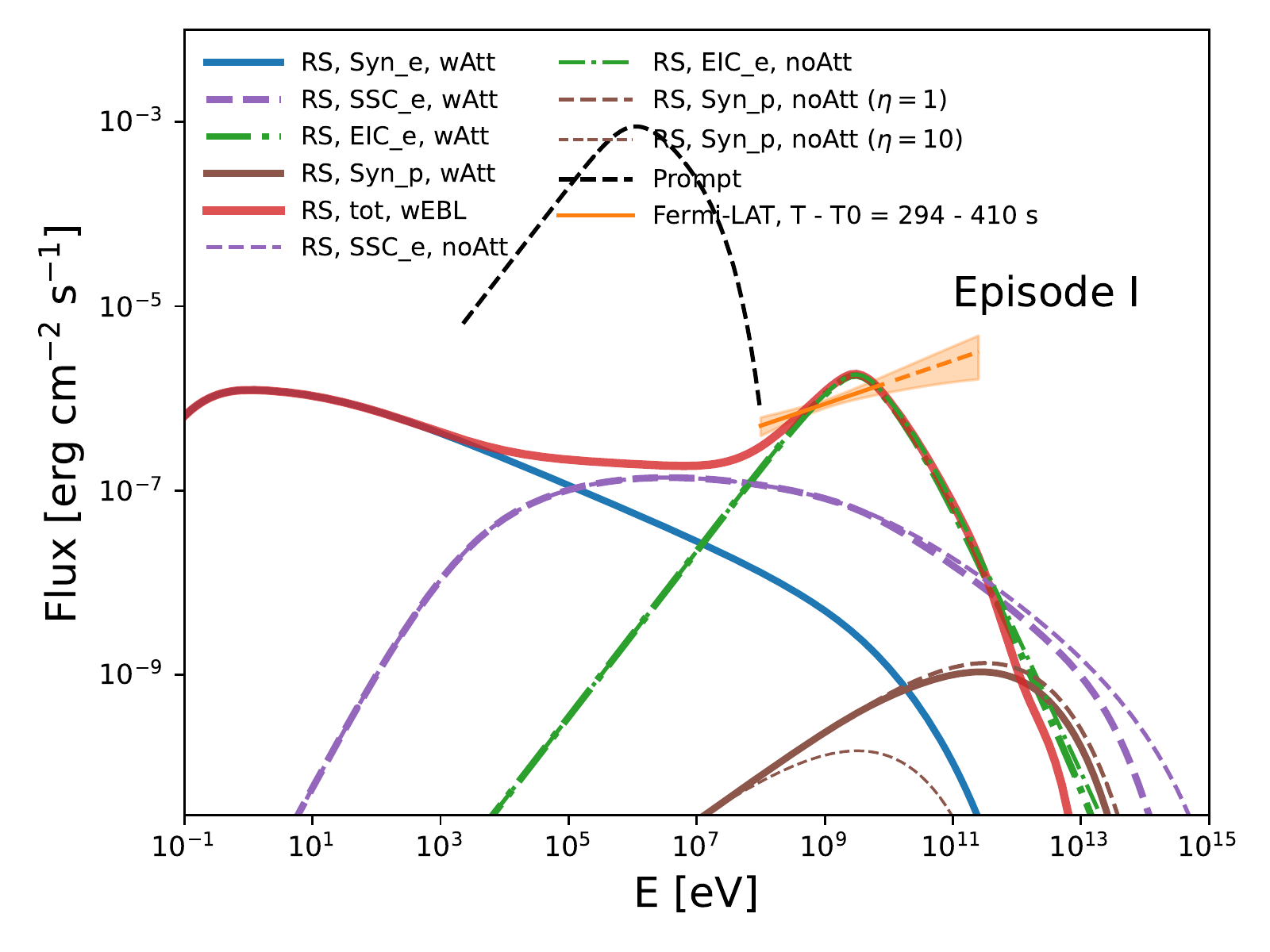}
\includegraphics[width=\linewidth]{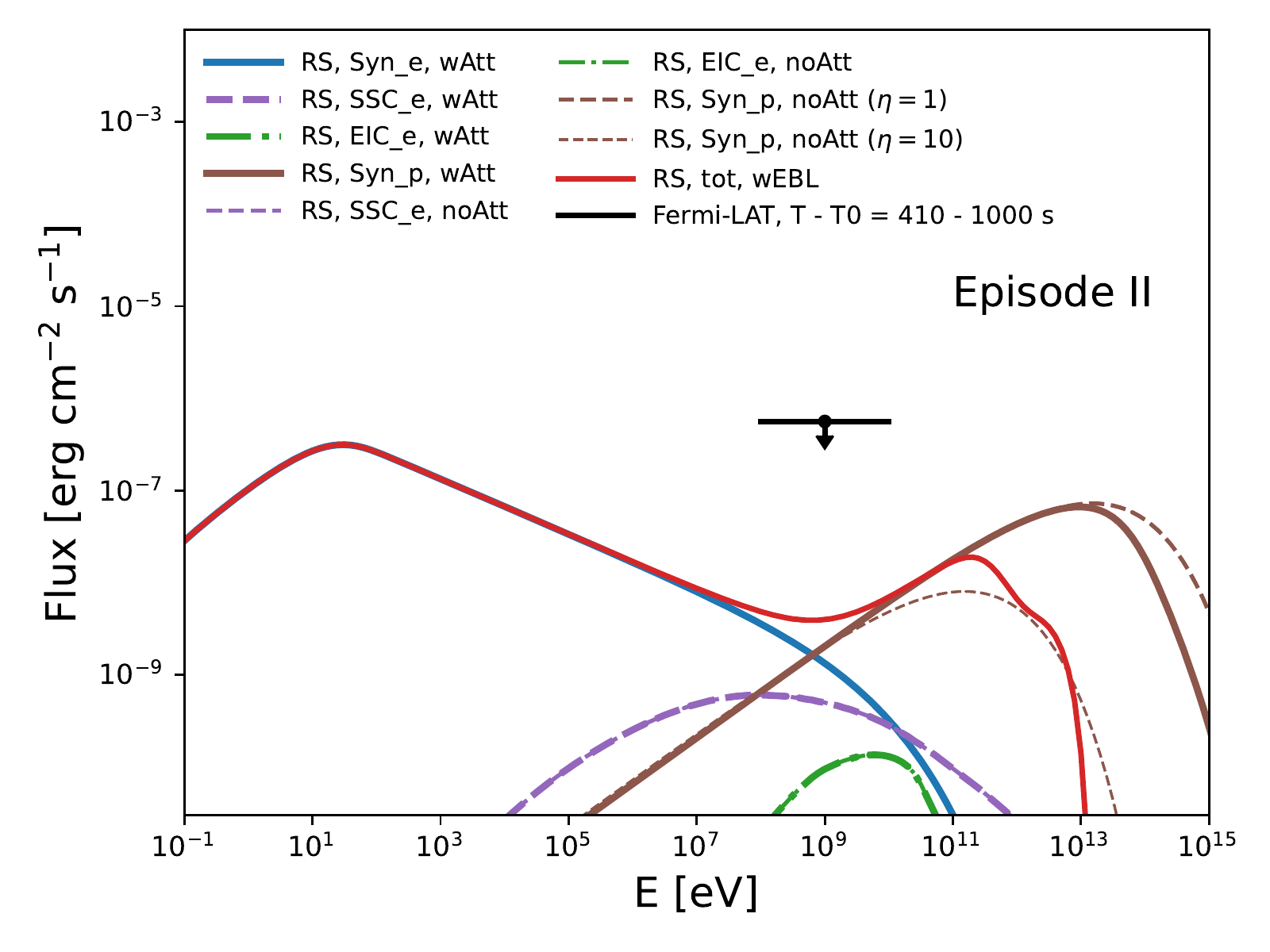}
\caption{Multi-wavelength energy spectra up to the VHE energy range in our reverse shock model in Episode I (upper panel) and Episode II (lower panel), respectively. 
The black dashed line in the upper panel is the prompt spectrum with an exponential cutoff at $E_{\gamma, \rm max} = 20\rm~MeV$. The orange line in the upper panel indicates the averaged energy spectrum observed by \textit{Fermi}-LAT within $T - T_0 = 294 - 410\rm~s$, extrapolated to the energy of 300 GeV. The black line in the low panel is the \textit{Fermi}-LAT 95\% UL within $T - T_0 = 410 - 1000\rm~s$.
}
\label{fig:RS-spectrum}
\end{figure}

The predicted multi-wavelength energy spectra for the two episodes are shown in Fig.~\ref{fig:RS-spectrum}, where the corresponding parameters are summarized in Table~\ref{table}.

\begin{table}
  \begin{threeparttable}
    \centering
	\caption{Physical parameters used in the reverse shock model.}
    \begin{tabular}{lccc}
    Parameter  & Episode I & Episode II \\
    \hline
    $\Gamma_0$ & 250 & 250 \\
    $\mathcal{E}_{k}$\tnote{a} [$\rm erg$] & $2 \times 10^{55}$ & $2 \times 10^{55}$ \\
    $n_{\rm ex}$ [$\rm cm^{-3}$] & 1 & 1 \\
    \hline
    $\delta T [$\rm s$]$ & ${\re 300} $\tnote{b} & ${\re 600}$ \\ 
    $\epsilon_B$ & 0.05 & 0.5 \\
    $\epsilon_e$ & 0.35 & 0.02 \\
    $f_e$  & 0.8& 0.01 \\
    $s_e$  & 2.6 & 2.6 \\
    $\epsilon_p$ & 0.1 & 0.08 \\
    $s_p$ & 2.0 & 2.0 \\
    $L_{\rm GRB\gamma}^{\rm iso}$ [$\rm erg~s^{-1}$] & $2 \times 10^{52}$ & $2 \times 10^{50}$ \\
    \hline
    \end{tabular}\label{table}
     \begin{tablenotes}
    \item [a] The dissipated energy during Episode I is only a fraction of the total kinetic energy $\mathcal{E}_{k, \rm I} \approx (\delta T_{\rm I} / \delta T) \mathcal{E}_k$ as explained in the main text. For Episode II, we have $\mathcal{E}_{k, \rm II} \approx \mathcal{E}_{k}$.
    \item [b] This is the duration of the GRB ejecta during Episode I $\delta T_{\rm I}$.
     \end{tablenotes}
  \end{threeparttable}
\end{table}

\subsection{Episode I - Upscattered prompt emission?}
In the upper panel of Fig.~\ref{fig:RS-spectrum}, we show the multi-wavelength energy spectrum from various processes for Episode I. 
The orange dashed region represents the average energy spectrum observed by \textit{Fermi}-LAT within $T - T_0 = 294 - 410\rm~s$\footnote{We do not consider gamma-rays detected in the first time interval $T - T_0 = 203 - 294\rm~s$ which overlaps with the LAT Bad Time Interval.}, extrapolated to 300 GeV. 
The dot-dashed curve is the upscattered prompt emission by the non-thermal electrons accelerated in the reverse shock region in Episode I which dominates energy flux beyond $\sim 1$ GeV, while the dashed curve is the SSC component, which dominates the energy flux from 0.1 GeV to 1 GeV. 
The synchrotron component is indicated as the blue solid curve, while the proton synchrotron component is marked as the brown solid curve. We also overlay the prompt emission assuming Band function with an exponential cutoff at $E_{\gamma, \rm max} = 20\rm~MeV$.
The corresponding parameters are summarized in Table.~\ref{table}.

At Episode I, the microphysical parameters satisfy $\epsilon_e > \epsilon_B$ and the inverse-Compton component dominates the gamma-ray emission.
We find that the EIC spectra can explain the hard energy spectrum observed by \textit{Fermi}-LAT above $\sim 1$ GeV.
Note that the detected $\sim 99.3\rm~GeV$ photon by \textit{Fermi}-LAT at $T - T_0 = 240\rm~s$ 
may be explained by the EIC component considering the enhancement of the prompt photons during the first time interval mentioned in Fig.~\ref{fig:lc}. 
However, the proton synchrotron component is not important in Episode I because the dominance of prompt MeV gamma-rays in the energy density limits the emission power of the synchrotron component.
We note that the predicted early optical flux could have been brighter than that of the naked-eye GRB 080319B~\citep{racusin_broadband_2008}.

\subsection{Episode II - Proton synchrotron emission?}
In the lower panel of Fig.~\ref{fig:RS-spectrum}, we show the energy spectrum from various processes for Episode II.
Due to the decrease of the prompt luminosity by at least two orders of magnitude, the EIC component is no longer important, while the proton synchrotron emission is more prominent at Episode II given that the magnetic fields are strong.
As shown in Table~\ref{table}, we assume that the microphysical parameter $\epsilon_B = 0.5$ is larger than the value adopted in Episode I. Such a change of $\epsilon_B$ is possible for long-lasting magnetic energy-dominated GRB ejecta.
Initially, the value of $\epsilon_B$ is very high for magnetic energy-dominated GRB ejecta. During Episode I, most of the GRB ejecta energy is released in the form of the extraordinarily strong prompt emission pulse with abundant pair production, which may effectively suppress $\epsilon_B$ especially if the prompt gamma-rays are produced by magnetic dissipation, and the reverse shock is stronger due to the weak magnetic energy in the GRB ejecta. Thus, we could expect strong inverse-Compton emission at Episode I. 
Later, the reverse shock continues to cross the inner region of the GRB ejecta, which is still magnetic energy dominated, and it becomes weak, consistent with the observations.
We can expect higher values of $\epsilon_B$ in Episode II, and the proton synchrotron emission is very efficient.
The above physical process could also explain the decrease of $\epsilon_e$ and $\epsilon_p$ in Episode II.

The peak energy from proton synchrotron emission can reach $\sim 10\rm~TeV$ without {\re extragalactic background light (EBL)} absorption.
Remarkably, the corresponding spectral index of the energy flux from proton synchrotron emission is $(3-s_p)/2 = 1/2$ for $s_p = 2$, which is larger than the spectral index {\re $\sim 0.2$} inferred from \textit{Fermi}-LAT observations in Episode I. 
The harder spectral index enhances the fraction of $\mathcal{O}(10\rm~TeV)$ photons in the total observed gamma-rays, even though it is undergoing EBL-induced attenuation during the propagation from the source to Earth~\citep[e.g.,][]{baktash_interpretation_2022, zhao_standard_2022}.

\subsection{High-energy neutrino production}
\begin{figure}[th]
\includegraphics[width=\linewidth]{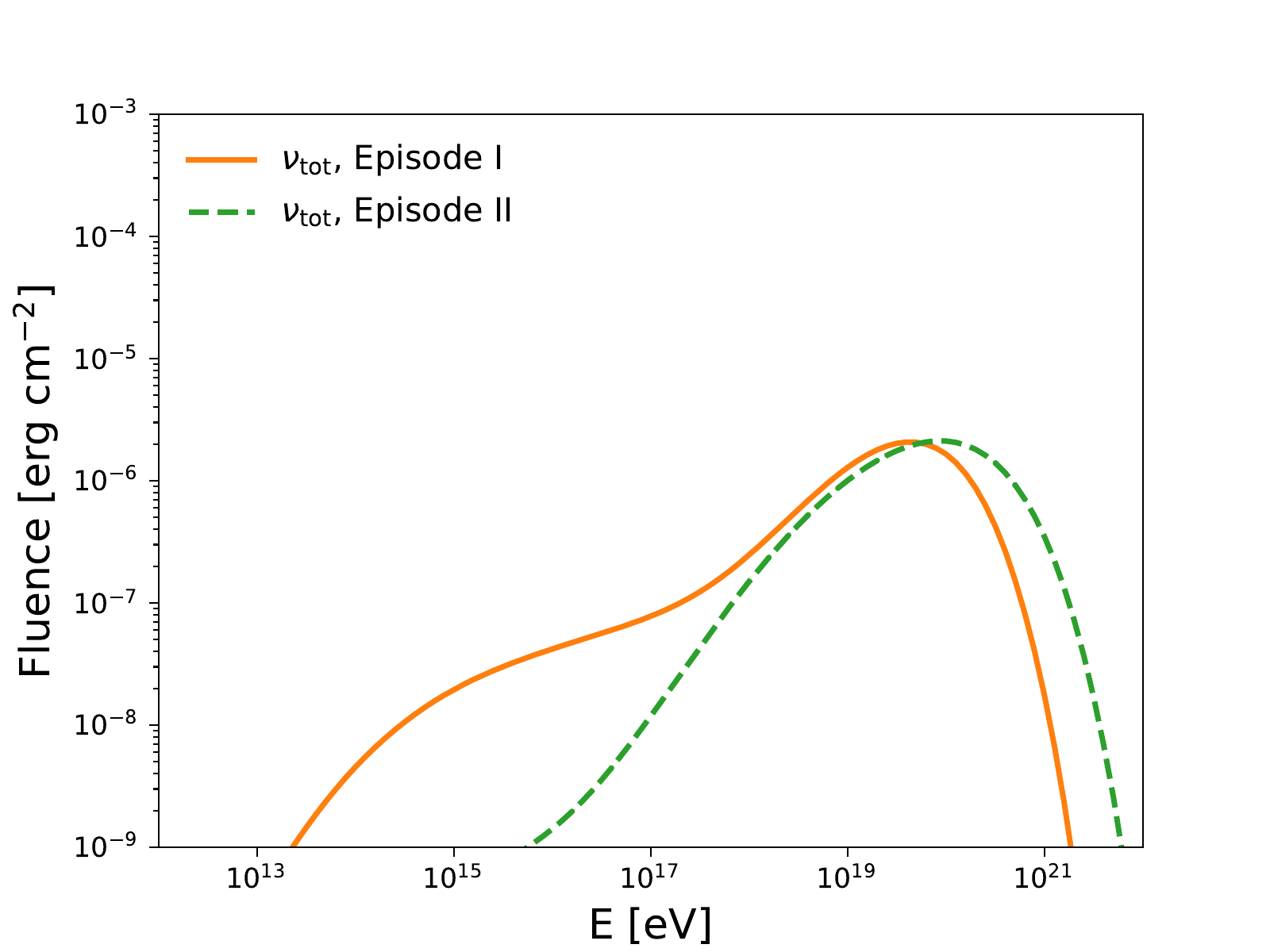}
\caption{Neutrino fluences emitted during the time window {\re $100\rm~s$} (Episode I) and {\re $300\rm~s$ (Episode II)}.}
\label{fig:RS-neutrino}
\end{figure}

In Fig.~\ref{fig:RS-neutrino}, we show the predicted neutrino fluences from Episode I and Episode II, respectively.
{\re The physical parameters used for the corresponding calculations are identical to those in Table~\ref{table}.}
The neutrino energy spectrum predicted in Episode I has two bumps, where the low-energy bump with peak energy at $\sim \rm~PeV$ energy range is due to the photomeson production interaction between high-energy protons with prompt target photon fields, while the high-energy bump with peak energy at $\sim 10\rm~EeV$ is the results between the interaction of UHE protons with the lower-energy synchrotron photons from the reverse shock.
The neutrino energy spectrum predicted in Episode II only shows one bump, which is consistent with the assumption that the effect of prompt emission is no longer important in Episode II.
The maximum proton energy is also higher in Episode II due to the strong magnetic fields, which is consistent with the neutrino spectrum where the neutrino has higher energy in Episode II.
Note that the cooling of secondary muons and pions is neglected when calculating the neutrino flux in our model, which will reduce the flux of the higher energy part of the observed neutrino spectrum, especially for Episode II~\citep{murase_high_2007}.
The predicted neutrino fluence is consistent with the upper limits reported by IceCube~\citep{the_icecube_collaboration_grb_2022, murase_neutrinos_2022, ai_model_2022}.

\section{Discussion and implications}\label{sec:dis}
\subsection{Anti-correlation}
The $\gamma \gamma$ optical depth $\tau_{\gamma\gamma}$ due to the interaction with prompt photons at the shocked ejecta region can be estimated with {\re the} following formula~\citep[e.g.,][]{murase_neutrinos_2022},
\begin{equation}
\tau_{\gamma\gamma}^{\rm prompt} \simeq {\re 11}\frac{\eta_{\gamma\gamma,-1}L_{\rm GRB\gamma, {\re 53.5}}^{\rm iso}}{r_{17.3}\Gamma_{\times, 2}^2 E_{\rm pk, \rm MeV}}
\begin{cases} (E_\gamma / \tilde{E}_{\gamma, b})^{\beta - 1}, E_\gamma < \tilde{E}_{\gamma, b} \\ (E_\gamma / \tilde{E}_{\gamma, b})^{\alpha - 1}, E_\gamma > \tilde{E}_{\gamma, b}
\end{cases},
\end{equation}
where $\eta_{\gamma\gamma} \sim 0.1$~\citep{svensson_non-thermal_1987}, $\tilde{E}_{\gamma, b} \approx \Gamma_\times^2 m_e^2 c^4 / E_{\rm pk, \rm MeV} \simeq 2.6~\Gamma_{\times, 2}^2 \rm~GeV$ {\re represents the typical energy of high-energy $\gamma$-rays that interact with target photons of peak energy $E_{\rm pk, MeV}$ in the observer frame}, and {\re $L_{\rm GRB\gamma}^{\rm iso}$ represents the isotropic-equivalent luminosity in the Konus-Wind band where the bolometric correction has been accounted for}.
The optical depth at 1 TeV is $\tau_{\gamma\gamma}(1\rm~TeV) \sim {\re 20}$ for $\alpha = 1.1$.
In addition, the synchrotron photons {\rm become} dominant in the energy range below $\lesssim \rm keV$ in Episode I compared to the prompt target photons as shown in Fig.~\ref{fig:RS-spectrum}. 
Then the optical depth can be estimated as 
\begin{equation}\label{eq:tau-syn}
\tau_{\gamma\gamma}^{\rm syn} \simeq 2300\frac{\eta_{\gamma\gamma,-1}L_{\gamma,50}^{\rm syn}}{r_{17.3}\Gamma_{\times, 2}^2 E_{\gamma, m}^{\rm syn}} 
\begin{cases}
 (E_\gamma / \tilde{E}_{\gamma, m})^{\frac{s_e}{2}}, E_\gamma < \tilde{E}_{\gamma, m} \\
 (E_\gamma / \tilde{E}_{\gamma, m})^{\frac{1}{2}}, \tilde{E}_{\gamma, c} > E_\gamma > \tilde{E}_{\gamma, m} 
 \end{cases}
\end{equation}
where $L_{\gamma}^{\rm syn}$ is the luminosity of synchrotron emission at
the synchrotron peak energy of the reverse shock component, $E_{\gamma, m}^{\rm syn} \simeq 1\rm~eV$ ($E_{\gamma, c}^{\rm syn} \simeq 0.1\rm~eV$) is the observed synchrotron peak (cooling) energy in the fast cooling regime, $\tilde{E}_{\gamma, m} \approx \Gamma_\times^2 m_e^2 c^4 / E_{\gamma, m}^{\rm syn} \simeq 2600~\Gamma_{\times, 2}^2 \rm~TeV$ and $\tilde{E}_{\gamma, c} \approx \Gamma_\times^2 m_e^2 c^4 / E_{\gamma, c}^{\rm syn} \simeq 26000~\Gamma_{\times, 2}^2 \rm~TeV$.
Then the optical depth at 10 TeV is $\tau_{\gamma\gamma}(10\rm~TeV) \sim 2$ for $s_e = 2.6$.
Note that the analytical calculation of the optical depth in Eq.~\ref{eq:tau-syn} is only valid for $ E_\gamma < \tilde{E}_{\gamma, c}$, {\re as the differential spectral index drops below unity ($2/3 < 1$) when the photon energy is below $E_{\gamma, c}^{\rm syn}$}.
Thus, in our model, we predict the anti-correlation of the $\gtrsim \rm TeV$ photons {\re during the strongest} prompt GRB emission phase, with the former escaping at later times.
For Episode II, due to the lower luminosity of prompt emission and the synchrotron photons, we can expect the escape of $\gtrsim \rm TeV$ photons simultaneously with low-energy photons.

\subsection{Implications for LHAASO detection}
\begin{figure}[th]
\includegraphics[width=\linewidth]{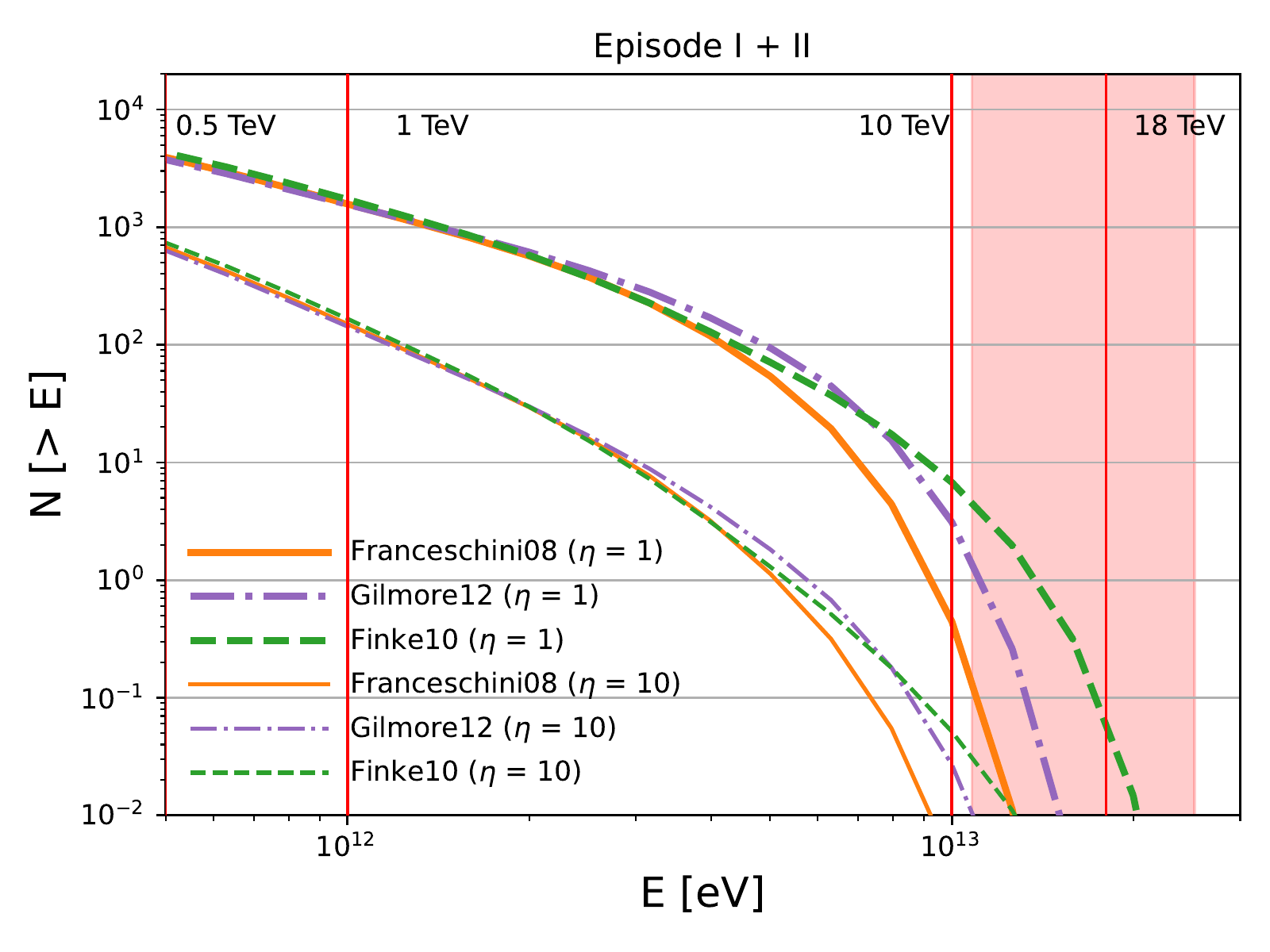}
\caption{The predicted detection (cumulative) number for the LHAASO detector with the energy spectrum predicted from our model using three different EBL models with {\re the total integration time $\Delta T=100$~s for Episode I and $\Delta T=300$~s for Episode II}. The vertical lines indicate the observed photon energy, and the shaded region {\re corresponds to the energy range $18 \times (100 \pm 40\%)\rm~TeV$ with a $\sim 40\%$ energy resolution at $18\rm~TeV$ energy band of LHAASO-KM2A~\citep{ma_chapter_2022}}.}
\label{fig:RS-detection}
\end{figure}

In Fig.~\ref{fig:RS-detection}, we show the number of photons predicted in our model that can be detected by LHAASO. 
Considering the uncertainties on the measured EBL  models, we adopt three representative EBL models, namely \textit{Franceschini08}~\citep{franceschini_extragalactic_2008}, \textit{Gilmore12}~\citep{gilmore_semi-analytic_2012} and \textit{Finke10}~\citep{finke_modeling_2010}.
We adopt the effective area of the LHAASO-WCDA for photons provided in~\cite{wang_chapter_2022} for zenith window $15^\circ < \theta < 30^\circ$ and the effective area from LHAASO-KM2A~\citep{ma_chapter_2022}.
The integrated number of events can be estimated as
\begin{equation}
N(>E) =\int_E^{E_{\rm max}} dE F(E) A_{\rm eff}^\gamma (E, \theta) \Delta T,
\end{equation}
where $F(E)$ is the predicted energy spectrum, $A_{\rm eff}^\gamma (E, \theta)$ is the effective area including both LHAASO-WCDA and LHAASO-KM2A, and $\Delta T$ is the integration time.
Even though the LHAASO-WCDA dominates the effective area at low energy, the effective area of LHAASO-KM2A is comparable to LHAASO-WCDA above $\sim 10\rm~TeV$~\citep{ma_chapter_2022}.

{\re If we adopt a typical integration time $300\rm~s$ for Episode II}, then the number of photons detected is $N(>0.5\rm~TeV) \sim {\re 4000}$ for photons with energy larger than 500 GeV, $N(>1\rm~TeV) \sim {\re 1500}$ for photons with energy larger than 1 TeV, and $N(>10\rm~TeV) \sim {\re 7}$ for photons with energy larger than 10 TeV.
The expected cosmic ray background for LHAASO-WCDA can be estimated with the same method used in~\cite{wang_chapter_2022}. We find that the photons with energies larger than 500 GeV (1 TeV) could be detected with LHAASO-WCDA with a significance level $\sim 25 \rm~s.d.$ ($\sim 15 \rm~s.d.$) estimated through Eq.~17 of ~\cite{li_analysis_1983} using fitted cosmic ray spectrum in~\cite{particle_data_group_review_2020}.

Even though we expect $N(>0.5\rm~TeV) \lesssim {\re 300}$ photons detected in Episode I, the detection of photons with energy above 1 TeV may be challenging for LHAASO-WCDA in Episode I, where the significance level is expected to be less than $\sim 2 \rm~s.d.$ in our fiducial case. 
TeV photons during Episode I mainly come from SSC emission in our model.
However, there is significant uncertainty in the SSC component, and if more $\gtrsim \rm TeV$ gamma-rays have been detected by LHAASO in Episode I, it would indicate that the SSC component becomes more important than the fiducial case shown in this work, {\re or the proton synchrotron emission is enhanced because of the larger value of $\epsilon_B$ as in the case of Episode II}.
{\re The VHE gamma-rays detected by LHAASO in Episode I may be primarily contributed by the SSC component resulting from non-thermal electrons that are accelerated by the forward shock~\citep[e.g.,][]{ren_very_2022, Sato:2022kup}. 
The fitting of the low-frequency radio to optical spectra at the earlier stages indicates that additional components, possibly resulting from synchrotron emission from ejecta swept by the reverse shock, are required beyond the standard forward shock model~\citep{OConnor:2023ieu}. 
However, further investigation using multi-wavelength data is needed to determine the relative contribution of the observed VHE gamma-rays from forward and reverse shock.
}

{\re In the thick shell case, it takes time for the shocked ejecta to get adjusted to the Blandford-McKee (BM) profile after the reverse shock crosses the ejecta. 
As in synchrotron emission from primary electrons accelerated at the reverse shock~\citep{kobayashiLightCurvesGamma2000}, we expect that the proton synchrotron light curve will decay with a temporal index of $-(73 s_p + 21)/96 \sim -1.7$ for $s_p = 2$, which is typically steeper than the forward shock case. 
However, the details of the reverse shock light curve depend on the profile of the ejecta, for long-lasting reverse shocks especially from $t_{\times,\rm I}$ to $t_{\times}$, which can be shallower than the standard case.}

The LHAASO-KM2A is extremely effective for suppressing the cosmic ray background~\citep{ma_chapter_2022}, so the detection of dozens of photons with energy $\mathcal{O}(10\rm~TeV)$ is reasonable considering the energy resolution.
The energy resolution at $\mathcal{O}(10\rm~TeV)$ is $\Delta E / E \sim 45\%$ for LHAASO-KM2A and $\Delta E / E \sim 60\%$ for LHAASO-WCDA. 
Our results indicate that even without considering the effect of new physics on the propagation of VHE gamma-rays, e.g., photon-ALP mixing~\citep{galanti_axion-like_2022, baktash_interpretation_2022, nakagawa_axion_2022, troitsky_parameters_2022} and Lorentz invariance violation (LIV)~\citep{li_lorentz_2022, zhu_light_2022, finke_possible_2022},
the detection of $\sim 18\rm~TeV$ photons by LHAASO-KM2A can be explained for reasonable EBL models.
Another possible explanation for the detection of $\sim 18\rm~TeV$ is the intergalactic electromagnetic cascade due to the propagation of UHECRs~\citep{batista_grb_2022, das_ultrahigh-energy_2022, mirabal_secondary_2022}. 
Both models also require the efficient production of UHECRs from GRBs. But in the intergalactic cascade scenario, the time delay of the photon arrival during the propagation of UHECRs and the development of the electromagnetic cascade depends strongly on the magnetic field structure of the host galaxy, the host galaxy cluster, and the intergalactic medium, and the time delay is longer given that the magnetic field strength is larger than $10^{-17}~\rm G$~\citep{Takahashi:2008pc, Murase:2008yy, Murase:2011cy, mirabal_secondary_2022}.

The default value of the coefficient {\re in the acceleration timescale formula $t_{\rm acc} = \eta t_L$ is set to $\eta = 1$, which may be optimistic.}
If we adopt a more considerable value of $\eta = 10$, then the proton synchrotron component would be reduced, which makes the detection of $\gtrsim \rm TeV$ gamma-rays more challenging in Episode II as shown in Fig.~\ref{fig:RS-detection}.
Thus, our results imply that the reverse shock of the long-lasting GRB 221009A should be a very efficient accelerator of non-thermal particles.
If most of the $\gtrsim \rm TeV$ gamma-rays are detected after Episode II, which is {\re $T \sim 600\rm~s$} in this work, then this would mean that the GRB ejecta duration is a bit longer and the reverse shock finishes crossing the GRB ejecta at later times.

\section{Summary}\label{sec:sum}
The detection of GRB 221009A in the VHE gamma-ray band up to $\mathcal{O}(10\rm~TeV)$ by LHAASO provides us with an opportunity to study the radiative processes of GRBs in the highest energy range.  
In this work, we studied the origin of VHE gamma-rays from GRB 221009A in the framework of the reverse shock model where non-thermal electrons and protons are expected to be accelerated.
The reverse shock could last up to thousands of seconds after the start of GRB, which is consistent with the long-lasting prompt emission observed from GRB 221009A.

We considered two episodes, where the emission from the reverse shock in the first episode (Episode I) is strongly affected by the strong prompt emission, while the effect of prompt emission becomes weak in the second episode (Episode II).  
In addition, the microphysical parameters, e.g., $\epsilon_B$ and $\epsilon_e$, could be different from Episode I to Episode II causing different behavior in the emission processes.

Our results show that the upscattered prompt MeV photons by the non-thermal electrons accelerated in the reverse shock region in Episode I mainly contribute to the energy flux observed by \textit{Fermi}-LAT above $\sim \rm GeV$, in addition to the SSC component.
The emission from the proton synchrotron process is not important in Episode I because of the presence of MeV gamma-rays and weak magnetic field strength.

We found that the proton synchrotron process can dominate the output in the VHE band in Episode II, where the magnetic field strength is strong enough to increase the proton synchrotron emission significantly.
Our rough estimates show that $\sim {\re 4000}$ photons with energy larger than 0.5 TeV can be detected by LHAASO in Episodes I and II, which may be consistent with the number of photons detected by LHAASO within 2000 seconds. 
Due to the hard spectral index of the proton synchrotron emission compared to the inverse-Compton process as inferred from \textit{Fermi}-LAT data, it is plausible to detect dozens of $\mathcal{O}(10\rm~TeV)$ photons using reasonable EBL models without invoking new physics.

Note that the forward shock has a similar energy to the reverse shock, and it may also contribute to the VHE emission.
The forward shock could enhance the number of $\sim\rm~TeV$ gamma-rays detected by LHAASO, but {\re difficult} for $\mathcal{O}(10\rm~TeV)$ gamma-rays, which emphasizes the role of proton synchrotron emission from the reverse shock as proposed in this work.

In the future, we can expect more GRBs to be detected in the VHE gamma-ray band, especially with the Cherenkov Telescope Array~\citep{inoue_gamma-ray_2013, kakuwa_prospects_2012}. Our work suggests that the observation of GRB in the VHE gamma-ray band can be used for constraining the particle acceleration and radiative processes of non-thermal electrons and protons in the reverse shock model.

\end{CJK*}

\begin{acknowledgements}
The work was partly supported by the NSF Grants No.~AST-1908689, No.~AST-2108466 and No.~AST-2108467, and KAKENHI No.~20H01901 and No.~20H05852 (K.M.) and No.~22H00130, 20H01901, 20H01904, 20H00158, 18H01215, 17H06357, 17H06362 (K.I.).

\end{acknowledgements}

\bibliographystyle{aasjournal}
\bibliography{ref}

\end{document}